\documentstyle[12pt,epsfig]{article}
\renewcommand{\bar}[1]{\overline{#1}}

\begin{document}

\begin{flushright}
GEF-Th-25/2006\\
\end{flushright}

\bigskip\bigskip
\begin{center}
{\large \bf ${\bf Q^2}$ Dependence of Azimuthal Asymmetries\\
in Semi-Inclusive Deep Inelastic Scattering\\
and in Drell-Yan}
\end{center}
\vspace{12pt}

\begin{center}
 {\bf Elvio Di Salvo\\}

 {Dipartimento di Fisica and I.N.F.N. - Sez. Genova, Via Dodecaneso, 33 \\-
 16146 Genova, Italy\\}
\end{center}

\vspace{10pt}
\begin{center} {\large \bf Abstract}

We study several azimuthal asymmetries in semi-inclusive deep inelastic scattering and in 
Drell-Yan, interpreting them within the formalism of the quark correlator, with a particular 
reference to T-odd functions. The correlator contains an undetermined energy scale, which we 
fix on the basis of a simple and rather general argument. We find a different value than the 
one assumed in previous treatments of T-odd functions. This implies different predictions on 
the $Q^2$ dependence of the above mentioned asymmetries. Our theoretical result on 
unpolarized Drell-Yan is compared with available data. Predictions on other azimuthal 
asymmetries could be tested against yields of planned experiments of Drell-Yan and 
semi-inclusive deep inelastic scattering.

\end{center}

\vspace{10pt}

\centerline{PACS numbers: 13.85.Qk, 13.88.+e}

\newpage
\section{Introduction}

Azimuthal asymmetries of remarkable size have been observed in various high energy inclusive 
reactions, especially in unpolarized Drell-Yan\cite{fa1,fa2,fa3}, in singly polarized 
semi-inclusive deep inelastic scattering[4-10] (SIDIS) and in inclusive production of 
(anti-)hyperons[11-15] and pions[16-19] from singly polarized hadronic collisions. The 
interpretation of such asymmetries from basic principles of QCD is quite challenging and has 
stimulated the interest of high energy physicists. In particular, in the present paper we 
focus our attention on the SIDIS and Drell-Yan asymmetries, which are somewhat analogous, 
since the two reactions are kinematically isomorphic. The theoretical activity about this 
subject is quite intense and lively, as witnessed by the numerous articles dedicated to the 
topic[20-70] in the last 15 years. 

An important element in the interpretation of such effects is the intrinsic transverse 
momentum of partons inside a hadron, whose crucial role in high energy reactions
has been  widely illustrated in the last years[23-25, 71-74]. Indeed, the transverse momentum 
is connected to the T-odd quark densities[20, 21, 27, 31-34], which provide a quite natural 
interpretation of the above mentioned asymmetries[27, 31-34, 42]. At the same time, the T-odd 
functions involve predictions of further azimuthal asymmetries in unpolarized and singly 
polarized inclusive reactions\cite{bjm,km}. 

These functions explain simultaneously\cite{bbh} the remarkable $cos 2\phi$ asymmetry and the 
negligible $cos \phi$ Fourier component exhibited by unpolarized Drell-Yan 
data\cite{fa1,fa2,fa3}, where $\phi$ is the usual azimuthal angle adopted in the 
phenomenological fits\cite{fa1,fa2,fa3}. The term $cos 2\phi$ may be just interpreted as a 
signature\cite{bbh} of the pair of chiral-odd (and T-odd) functions involved in this picture.  
However, the current treatment of the T-odd functions does not reproduce the dependence of  
this asymmetry on the effective mass of the Drell-Yan lepton pair.
More generally, some doubts have been cast on the $Q^2$ dependence of the transverse momentum 
distribution functions\cite{te1,te2,dis1,mw}, where $Q$ is the QCD hard scale. This imposes a 
revision of the parameterization of the transverse momentum quark correlator, a fundamental 
theoretical tool for cross section calculations at high energies. This quantity - originally 
introduced by Ralston and Soper in 1979\cite{rs} and successively improved by Mulders and 
Tangerman\cite{mt,tm1,tm2} (see also the more recent contributions on the 
subject\cite{bam,gm1,gm2}) - consists of a $4\times 4$ matrix. Therefore it may be 
parameterized according to the components of the Dirac algebra, taking into account the 
available vectors and hermiticity and Lorentz and parity invariance. The parameterization - 
whose coefficients are the quark distribution functions inside the hadron - includes an 
undetermined energy scale, $\mu_0$ \cite{ko}, usually assumed\cite{tm1,tm2,mt} equal to the 
mass of the hadron related to the active quark. We shall see that this choice is not unique, 
perhaps not the most appropriate in normalizing some "leading twist" functions. 
Alternatively, we propose $\mu_0 = Q/2$, which explains quite naturally the $Q^2$ dependence 
of the unpolarized Drell-Yan asymmetry. Moreover, concerning the SIDIS and other Drell-Yan 
azimuthal asymmetries, we get predictions which contrast with those given by previous 
authors, and which could be tested against present\cite{her1,her2,her3,her4,co}, 
forthcoming\cite{cls} and future\cite{rh,gs1,gs2,fn} data.

Here we shall not study all azimuthal asymmetries considered in the 
literature\cite{om1,om2,om3,msc}, we shall limit ourselves to SIDIS of unpolarized or 
longitudinally polarized lepton beams off unpolarized or transversely polarized targets, and 
to unpolarized or singly polarized  Drell-Yan, with transverse polarization; moreover, we 
shall consider just the asymmetries usually classified as leading twist\cite{mt,bm,bjm}.

The paper is organized as follows. In sect. 2 we give the general formulae for the SIDIS and 
Drell-Yan cross sections, introducing the formalism of the correlator; in particular we 
illustrate in detail the T-odd functions. Sect. 3 is dedicated to the theoretical formulae 
for azimuthal asymmetries. In sect. 4 we determine the parameter $\mu_0$, by comparing the 
correlator with the quark density matrix in QCD parton model. Such a determination leads to 
predictions on the $Q^2$ dependence of the asymmetries, which we illustrate in sect. 5. In 
sect. 6 we compare our results with experimental data, as regards unpolarized Drell-Yan. 
Lastly we draw a short conclusion in sect. 7.

\section{SIDIS and Drell-Yan Cross Sections}
\subsection{General formulae}
Consider the SIDIS and the Drell-Yan reactions, {\it i. e.},
\begin{equation}
l h_A \to l' h_B X ~~~~~~ {\mathrm and} ~~~~~~ h_A h_B \to l^+ l^- X,\label{rr}
\end{equation}
where the $l$'s are charged leptons and the $h$'s are hadrons. Incidentally, these two 
reactions are topologically equivalent\cite{co1}. At not too high energies one can adopt 
one-photon exchange approximation, where the cross sections for such reactions have an 
expression of the type
\begin{equation}
\frac{d\sigma}{d\Gamma} =
\frac{(4\pi\alpha)^2}{4{\cal F}Q^4} L^{\mu\nu} W_{\mu\nu}.  \label{thecs}
\end{equation}
Here $d\Gamma$ is the phase space element, $\alpha$ the fine structure constant and ~~ ${\cal 
F}$ ~~ $=$ ~~ $\sqrt{(p_1\cdot p_2)^2-m_1^2m_2^2}$ ~~ the flux factor, $p_i$ and $m_i$ ($i = 
1,2$) being the 4-momenta and the masses of the initial particles. Moreover
$L^{\mu\nu}$ and $W^{\mu\nu}$ are respectively the leptonic and hadronic tensor. In 
particular, we have
\begin{equation}
L^{\mu\nu} = \ell^{\mu} {\bar \ell}^{\nu} + \ell^{\nu} {\bar \ell}^{\mu} - 
g^{\mu\nu} \ell \cdot {\bar \ell}, \label{lts}
\end{equation}
where $\ell$ and ${\bar \ell}$ are the four-momenta of the initial and final lepton (in 
SIDIS) or of the two final leptons (in Drell-Yan). As regards the hadronic tensor, one often 
adopts, in the framework of the factorization theorem\cite{co3,co4,co1,co2}, the so-called 
"handbag" approximation, where all information concerning the "soft" functions of the quark 
inside the hadrons is encoded in a parameterization of the quark-quark correlator, according 
to the various Dirac components\cite{tm1,mt}. In this approximation the hadronic tensor reads
\begin{equation}
W^{\mu\nu} = c\sum_ae^2_a\int d^2p_{\perp} Tr\left[\Phi^a_A(x_a, {\bf p}_{\perp}) 
\gamma^{\mu}\Phi^b_B(x_b, {\bf q}_{\perp}-{\bf p}_{\perp}) \gamma^{\nu}\right]. \label{ht}
\end{equation}
Here $c$ is due to color degree of freedom, $c = 1$ for SIDIS and $c = 1/3$ for Drell-Yan.
$\Phi_A$ and $\Phi_B$ are correlators, relating the active (anti-)quarks  to the (initial or 
final) hadrons $h_A$ and $h_B$. $a$ and $b$ are the flavors of the active partons, with $a = 
u,d,s,\bar{u},\bar{d},\bar{s}$ and $b = a$ in SIDIS, $b = \bar{a}$ in Drell-Yan; $e_a$ is the 
fractional charge of flavor $a$. In Drell-Yan $\Phi_A$ and $\Phi_B$ encode information on 
(anti-)quark distributions inside the initial hadrons: the $x'$s are the longitudinal 
fractional momenta of the active quark and antiquark, ${\bf p}_{\perp}$ is the transverse 
momentum of the active parton of $h_A$ and ${\bf q}_{\perp}$ is the transverse momentum of 
the lepton pair. In SIDIS $\Phi_B$ is replaced by the fragmentation correlator $\Delta[z, 
z({\bf q}_{\perp}-{\bf p}_{\perp})]$, describing the fragmentation of the struck quark into 
the final hadron $h_B$ (see subsect. 2.4). Here $z$ is the longitudinal fractional momentum 
of $h_B$ with respect to the fragmenting quark and $z{\bf q}_{\perp}$ is the transverse 
momentum of $h_B$ with respect to the virtual photon momentum. Approximation (\ref{ht}) holds 
for the hadronic tensor under the condition\cite{bbh,bhs1,bhs2,bhs3} 
\begin{equation}
q_{\perp} << Q, \label{ineq}
\end{equation}
where $q_{\perp} = |{\bf q}_{\perp}|$. Moreover we neglect the Sudakov 
suppression\cite{bo0,bo1}, as usually assumed at moderate $Q^2$[42, 54-62]. 

\subsection{Parameterization of the Correlator}

The correlator for a nucleon may be parameterized according to the Dirac algebra, taking into 
account hermiticity and Lorentz and parity invariance. It is conveniently split into a T-even 
and a T-odd term, {\it i. e.}, 
\begin{equation}
\Phi = \Phi_e + \Phi_o, \label{eo}
\end{equation}
where $\Phi_{e}$ is even under time reversal and $\Phi_{o}$ is odd under the same 
transformation. At leading twist one has\cite{mt,bhm,gm1,gm2} 
\begin{eqnarray}
\Phi_e &\simeq& \frac{\cal P}{\sqrt{2}}  \left\{f_1\rlap/n_+ + (\lambda 
g_{1L}+\lambda_{\perp}g_{1T})\gamma_5\rlap/n_+ + 
\frac{1}{2}h_{1T}\gamma_5[\rlap/S_{\perp},\rlap/n_+]\right\}\nonumber
\\ 
&+& \frac{\cal P}{2\sqrt{2}}\left(\lambda h^{\perp}_{1L}+\lambda_{\perp} 
h^{\perp}_{1T}\right) \gamma_5 [\rlap/\eta_{\perp},\rlap/n_+],\label{par0} 
\\
\Phi_o &\simeq& \frac{\cal P}{\sqrt{2}}\left\{f^{\perp}_{1T} 
\epsilon_{\mu\nu\rho\sigma}\gamma^{\mu}n_+^{\nu}\eta_{\perp}^{\rho}S_{\perp}^{\sigma}+
ih^{\perp}_1\frac{1}{2}[\rlap/\eta_{\perp},\rlap/n_+]\right\}. \label{cmo}
\end{eqnarray}
In formulae (\ref{par0}) and (\ref{cmo}) we have used the notations of refs.\cite{mt,tm1} for 
the "soft" functions\footnote{The correlator (\ref{eo}) has a different normalization than in 
ref.\cite{mt}, in accord with the definition (\ref{ht}) of the hadronic tensor.}. $n_{\pm}$ 
are lightlike, dimensionless vectors, such that $n_+\cdot n_- = 1$ and whose space components 
are along (+) or opposite to (-) the nucleon momentum. Moreover  the Pauli-Lubanski vector of 
the nucleon, denoted as $S$ and such that $S^2 = -1$, may be decomposed as
\begin{equation}
S =  \lambda \frac{P}{M} +S_{\perp}\label{pl}.
\end{equation}
Here $P$ is the nucleon four-momentum, with $P^2 = M^2$,  $\lambda = -S\cdot n_0$ and $n_0 
\equiv (0,0,0,1)$ in the nucleon rest frame, taking the $z$-axis along the nucleon momentum. 
Thirdly,  
\begin{eqnarray}
{\cal P} &=& \frac{1}{\sqrt{2}}p\cdot n_-, \ ~~~~~ \ ~~~~~ \ \lambda_{\perp} = -S\cdot 
\eta_{\perp},\label{li}
\\
\eta_{\perp} &=& p_{\perp}/\mu_0, \ ~~~~~ \ p_{\perp} = p - (p\cdot n_-)n_+ - (p\cdot 
n_+)n_-\label{ms}
\end{eqnarray}
and $p$ is the quark four-momentum. Notice that the parameter ${\cal P}$ is similar to the 
one introduced by Jaffe and Ji\cite{jj1,jj2} with the same notation: denoting that parameter 
by
${\cal P}_{JJ}$, one has ${\cal P} = x{\cal P}_{JJ}/\sqrt{2}$.
Lastly, the energy scale $\mu_0$, encoded in the dimensionless vector $\eta_{\perp}$, has 
been introduced in such a way that all functions involved in the parameterization of $\Phi$ 
have the dimensions of a probability density. This scale - defined for the first time in 
ref.\cite{ko}, where it was denoted by $m_D$ - determines the normalization of the functions 
which depend on $\eta_{\perp}$; therefore $\mu_0$ has to be chosen in such a way that these 
functions may be interpreted just as probability densities. We shall see in sect. 4 that 
taking $\mu_0$ equal to the rest mass of the hadron, as usually done\cite{tm1,tm2,mt}, is 
not, perhaps, the most appropriate in this sense. Two observations are in order about 
$\mu_0$. First of all, it is washed out by integration over ${\bf p}_{\perp}$ of the 
correlator, therefore it does not influence the common\cite{jj1,jj2} distribution functions. 
Secondly, we can reasonably assume that, for sufficiently large $Q^2$, this parameter is 
independent of the perturbative interactions among partons: as we shall see, this conjecture 
can be proved rigorously.  

\subsection{T-odd functions}

As explained in the introduction, the T-odd functions deserve especial attention. In 
particular, the two functions introduced in formula (\ref{cmo}) may be interpreted as quark 
densities: $h_1^{\perp}$ corresponds to the quark transversity in an unpolarized (or 
spinless) hadron, while $f^{\perp}_{1T}$ is the density of unpolarized quarks inside a 
transversely polarized spinning hadron\cite{bm,dis1}. 

A possible mechanism for generating these effects has been analyzed in detail, from different 
points of view, by various authors[35-41, 44-53]. In particular, the function 
$f^{\perp}_{1T}$, known as the Sivers function, may give rise to a single spin  asymmetry,  
as predicted for the first time many years ago by Sivers\cite{si1,si2} as a consequence of 
coherence among partons. Essential ingredients for producing the effect 
are\cite{bhs1,bhs2,bhs3} 

a) two amplitudes with different quark helicities and different components ($\Delta L_z = 1$) 
of the orbital angular momentum;

b) a phase difference between such amplitudes, caused, for example, by one gluon exchange 
between the spectator partons and the active quark, either before or after the hard 
scattering: owing to the different orbital angular momenta, the gluon interaction causes a 
different phase shift in the two amplitudes. 

Incidentally, a $\Delta L_z = 1$ is connected to the anomalous magnetic moment of the 
nucleon[35-37, 48-53]; however, the difference in quark helicities could be attributed, in 
part, also to spontaneous chiral symmetry breaking\cite{co2}. In this connection, we think 
that the correct origin and the basic mechanisms for producing the Sivers asymmetry should be 
investigated more deeply.

The initial and final state interactions may be described by the so-called link operator, 
introduced in the definition of bilocal functions in order to assure gauge 
invariance\cite{mt,co2,jy1,jy2,jy3}. Moreover they cause also a nonvanishing $h_1^{\perp}$ 
\cite{gg1,gg2,gg3,gg4}: in a scalar diquark model, this function is equal to $f_{1T}^{\perp}$ 
\cite{bbh} (see also ref.\cite{gg4}). In the mechanism which generates quark transversity in 
an unpolarized nucleon, angular momentum conservation implies a change by one or two units of 
orbital angular momentum of the quark; this change can be connected to a pseudovector 
particle exchange, while the above mentioned initial or final state interactions are 
interpreted as Regge (or absorptive) cuts\cite{gg1,gg2,gg3,gg4}.    
  
From the above discussion it follows that quark-gluon interactions are essential for 
producing T-odd functions. Indeed, if such interactions are turned off, T-odd functions are 
forbidden by time reversal invariance\cite{co1} in transverse momentum space. On the 
contrary, they are allowed in the impact parameter space[48-53]: the Sivers asymmetry can be 
viewed as a left-right asymmetry with respect to the nucleon spin in that space, where final 
state interactions produce a chromodynamic lensing for the struck quark[48-53].

As shown in appendix, T-odd functions can be related\cite{bmp,mw} to the Qiu-Sterman 
\cite{qs1,qs2,qs3} effect, which takes into account quark-gluon-quark correlations (see also 
ref.\cite{et2,et3}). This relation will be illustrated especially at the end of sect. 5, in 
connection with singly polarized Drell-Yan: T-odd functions turn out to produce an asymmetry 
whose $Q^2$ dependence coincides with the one obtained by assuming one of the above mentioned 
correlations\cite{bmt,dis4}. T-odd functions can be approximately factorized\cite{co2} - up 
to a sign, according as to whether the functions are involved in SIDIS or in 
Drell-Yan\cite{co2,bmp} - if condition (\ref{ineq}) is fulfilled\cite{bbh}; otherwise one is 
faced with serious difficulties as regards universality of the effect\cite{pe}.

\subsection{Fragmentation Correlator}

The fragmentation correlator can be parameterized analogously to $\Phi$, see subsect. 2.2. We 
have, in the case of quark fragmentation into a pion, 
\begin{equation}
\Delta = \Delta_e+\Delta_o,\label{frg}
\end{equation}
where, at leading twist, the T-even part is given by
\begin{equation}
\Delta_e \simeq \frac{1}{2} k\cdot n'_+ D \rlap/n'_-
\end{equation}
and the T-odd part reads
\begin{equation}
\Delta_o \simeq \frac{1}{4\mu_0^{\pi}}H_1^{\perp} [\rlap/k_{\perp},\rlap/n'_-]. \label{fc}
\end{equation}
Here $k$ is the four-momentum of the quark, $k_{\perp} = k - (k\cdot n'_-)n'_+ - (k\cdot 
n'_+)n'_-$ and $n'_\pm$ are a pair of lightlike vectors, defined analogously to $n_\pm$, but 
such that the space component of $n'_-$ is along the pion momentum. $\mu_0^{\pi}$ is an 
energy scale analogous to $\mu_0$. Lastly $D$ and $H_1^{\perp}$ are fragmentation functions, 
$D$ is the usual one, chiral even, while $H_1^{\perp}$  - the Collins function\cite{co1} - is 
chiral odd.  It is important to notice that the latter function is interaction dependent, as 
well as the T-odd distribution functions: indeed, it has been shown\cite{le} that this 
function would vanish in absence of interactions among partons. 
 
\section{SIDIS and Drell-Yan Asymmetries}

Now we deduce the expressions of the asymmetries involved in the two reactions considered, 
according to the formalism introduced in the previous section (see also, {\it e. g.,} 
refs.\cite{mt,bjm,ds,bd} for SIDIS and ref.\cite{bo} for Drell-Yan). As regards SIDIS, we 
treat the cases where the initial lepton is either unpolarized or longitudinally polarized, 
while the nucleon target is either unpolarized or tranversely polarized. On the other hand, 
concerning Drell-Yan, we consider the situations where at most one of the two initial hadrons 
(typically a proton) is transversely polarized. For our aims, the most relevant kinematic 
variables are two azimuthal angles, denoted as $\phi$ and $\phi_S$. In the case of Drell-Yan 
they are the azimuthal angles, respectively, of the momentum of the positive lepton and of 
the spin of the initial polarized hadron, in a frame at rest in the center of mass of the 
final lepton pair.
Different frames, related to one another by rotations, have been defined: while the $x$-axis 
is usually taken along ${\bf q}_{\perp}$, the $z$-axis may be taken along the beam momentum - 
Gottfried-Jackson(GJ) frame\cite{gj} -, antiparallel to the target momentum - U-channel (UC) 
frame\cite{fa1,fa2,fa3} -, or along the bisector of the beam momentum and of the direction 
opposite to the target momentum - Collins-Soper (CS) frame\cite{cs}. We shall discuss the 
frame dependence of the asymmetry parameters later on. As far as SIDIS is concerned, $\phi$ 
and $\phi_S$ are respectively the azimuthal angles - defined in the Breit frame where the 
proton momentum is opposite to the photon momentum - of the final hadron momentum and of the 
target spin vector with respect to the production plane.

\subsection{SIDIS Asymmetries}

The doubly polarized SIDIS cross section, with a longitudinally polarized lepton and a 
tranversely polarized nucleon, may be written as a sum of 4 terms, {\it i. e.},
\begin{equation}
\left(\frac{d\sigma}{d\Gamma}\right) = \left(\frac{d\sigma}{d\Gamma}\right)_{UU} + 
\left(\frac{d\sigma}{d\Gamma}\right)_{UT} + \left(\frac{d\sigma}{d\Gamma}\right)_{LU} 
+\left(\frac{d\sigma}{d\Gamma}\right)_{LT}.
\end{equation}
Here we have singled out the unpolarized $(UU)$, the singly polarized - either with a 
transversely polarized target, $(UT)$, or with a longitudinally polarized beam, $(LU)$ - and 
the doubly polarized $(LT)$ contributions. According to the formalism introduced in sect. 2, 
we get, at leading twist approximation,
\begin{eqnarray}
\left(\frac{d\sigma}{d\Gamma}\right)_{UU} &\simeq& \sum_ae_a^2[U^a_0 + U^a_1 cos2\phi],
\label{unp}
\\
\left(\frac{d\sigma}{d\Gamma}\right)_{UT} &\simeq& \sum_ae_a^2[S^a_1 sin(\phi+\phi_S)+ S^a_2 
sin(\phi-\phi_S)
\nonumber
\\
&+& S^a_3 sin(3\phi-\phi_S) +S^a_4 sin2\phi], \label{ssa}
\\
\left(\frac{d\sigma}{d\Gamma}\right)_{LU} &\simeq& 0,
\\
\left(\frac{d\sigma}{d\Gamma}\right)_{LT} &\simeq& \sum_ae_a^2[D^a_1 + D^a_2 
cos(\phi-\phi_S)]. \label{dsa}
\end{eqnarray}
Here we have expressed the cross section in units $\alpha^2xz^2s/Q^4$, where $s$ is the 
overall c.m. energy squared. Moreover, omitting the flavor indices of the functions involved, 
we have
\begin{eqnarray}
U_0 &=& A(y){\cal F}[w_{U_0}, f_1, D], \label{u0}
\\
U_1 &=& -C(y) \frac{q_{\perp}^2}{\mu_0\mu_0^{\pi}}{\cal F}[w_{U_1}, h_1^{\perp}, 
H_1^{\perp}], \label{u2} 
\\
S_1 &=& C(y) |{\bf S}_{\perp}|\frac{q_{\perp}}{\mu_0^{\pi}}{\cal F}[w_{S_1}, h_{1T}, 
H_1^{\perp}], 
\\ 
S_2 &=& A(y) |{\bf S}_{\perp}|\frac{q_{\perp}}{\mu_0}{\cal F}[w_{S_2}, f_{1T}^{\perp}, D],
\\
S_3 &=& C(y) |{\bf S}_{\perp}|\frac{q_{\perp}^3}{\mu_0^2\mu_0^{\pi}}{\cal F}[w_{S_3}, 
h_{1T}^{\perp}, H_1^{\perp}],
\\ 
S_4 &=& \lambda C(y)\frac{q_{\perp}^2}{\mu_0\mu_0^{\pi}}{\cal F}[w_{S_4}, h_{1L}^{\perp}, 
H_1^{\perp}],
\\
D_1 &=& \lambda \lambda_{\ell} \frac{1}{2} E(y){\cal F}[w_{D_1}, g_{1L}, D], 
\\ 
D_2 &=&  \lambda_{\ell} \frac{1}{2} E(y) \frac{q_{\perp}}{\mu_0^{\pi}}{\cal F}[w_{D_2}, 
g_{1T}, D]. \label{d1}
\end{eqnarray}
We have denoted by $\lambda_{\ell}$ and ${\bf S}_{\perp}$ respectively the helicity of the 
initial lepton and the transverse component of the nucleon spin vector, with $|{\bf 
S}_{\perp}| = sin\phi_S$ and $\lambda = cos\phi_S$. Moreover 
\begin{equation} 
A(y) = 1-y+1/2 y^2,  \ ~~~~~ \ C(y) = 1-y, \ ~~~~~ \ E(y) = y(2-y),
\end{equation} 
where $y \simeq q^-/\ell^-$ and $q$ is the four-momentum of the virtual photon, such that 
$|q^2| = Q^2$. Lastly, ${\cal F}$ is a functional\cite{bjm}, 
\begin{equation} 
{\cal F}[w, f, D] = \int d^2p_{\perp} w({\bf p}_{\perp},{\bf q}_{\perp}) f({\bf p}_{\perp})
D[z, z({\bf q}_{\perp}-{\bf p}_{\perp})],
\end{equation} 
$w$, $f$ and $D$ being, respectively, a weight function, a distribution function and a 
fragmentation function. As to the weight functions, we have
\begin{eqnarray}
w_{U_0} &=& w_{D_1} = 1,\label{ws0}
\\ 
w_{U_1} &=& w_{S_4} = 2{\hat{\bf u}}\cdot {\hat{\bf p}}_{\perp} {\hat{\bf u}}\cdot {\hat{\bf 
k}_{\perp}} - {\hat{\bf k}}_{\perp} \cdot {\hat{\bf p}}_{\perp}, \label{ws1}
\\
w_{S_1} &=&  {\hat{\bf u}}\cdot {\hat{\bf k}}_{\perp}, \ ~~~~~~~ \ w_{S_2} = w_{D_2} = 
{\hat{\bf u}}\cdot {\hat{\bf p}}_{\perp},
\\
w_{S_3} &=& 4({\hat{\bf u}}\cdot {\hat{\bf p}}_{\perp})^2 {\hat{\bf u}}\cdot {\hat{\bf 
k}}_{\perp} - 2{\hat{\bf u}}\cdot {\hat{\bf p}}_{\perp}{\hat{\bf k}}_{\perp} \cdot {\hat{\bf 
p}_{\perp}} - {\hat{\bf u}}\cdot {\hat{\bf k}}_{\perp}{\hat{\bf p}}_{\perp}^2. \label{ws5}
\end{eqnarray}
Here we have set ${\hat{\bf u}} = {\bf q}_{\perp}/q_{\perp}$, ${\hat{\bf p}}_{\perp} = {\bf 
p}_{\perp}/q_{\perp}$ and ${\hat{\bf k}}_{\perp}$ = $({\bf q}_{\perp}-{\bf 
p}_{\perp})/q_{\perp}$. Notice that the first two terms of the cross section (\ref{ssa}) 
correspond respectively to the Collins and Sivers asymmetry\cite{co1,si1,si2}. 

\subsection{Weighted Asymmetries in SIDIS}

The weighted asymmetries are defined as
\begin{equation} 
A_W = \frac{\langle W\rangle}{\langle 1\rangle}.
\end{equation} 
Here brackets denote integration of the weighted cross section over $q_{\perp}$ and over the 
azimuthal angles defined above.  $W$ is the weight function, consisting of the Fourier 
component we want to pick up [see eqs.
(\ref{unp}) to (\ref{dsa})], times $(q_{\perp}/\mu_0)^{n_a}(q_{\perp}/\mu_0^{\pi})^{n_b}$, 
where $n_a$ and $n_b$ are respectively the powers with which $\hat{{\bf p}}_{\perp}$ and 
$\hat{{\bf k}}_{\perp}$ appear in the functions $w$ [see eqs. (\ref{ws0}) to (\ref{ws5})].  
For instance, the weight function corresponding to the Collins asymmetry is $W_{S_1} = 
(q_{\perp}/\mu_0^{\pi})sin(\phi+\phi_S)$. 

\subsection{Drell-Yan Asymmetries}

In the case of singly polarized Drell-Yan with a transversely polarized proton, we have (see 
also ref.\cite{bo}) 
\begin{eqnarray}
\left(\frac{d\sigma}{d\Gamma'}\right)_{UU} &=& \sum_ae_a^2[U^{'a}_0 + U^{'a}_1 cos 2\phi], 
\label{dyu}
\\
\left(\frac{d\sigma}{d\Gamma'}\right)_{UT} &=& \sum_ae_a^2[S^{'a}_1 sin(\phi+\phi_S)+ 
S^{'a}_2 sin(\phi-\phi_S)
\nonumber
\\
&+& S^{'a}_3 sin(3\phi-\phi_S)]. \label{dys}
\end{eqnarray}
Here we have adopted the same approximation as before and have expressed the cross section in 
units $\alpha^2/3Q^2$. Moreover
\begin{eqnarray}
U'_0 &=& A'(y){\cal F}[w_{U_0}, f_1, {\bar f}_1],\label{u0p} 
\\ 
U'_1 &=& C'(y)\frac{q_{\perp}^2}{\mu_0\mu'_0}{\cal F}[w_{U_1}, h_1^{\perp}, {\bar 
h}_1^{\perp}], \label{u1p}
\\
S'_1 &=& -C'(y)\frac{q_{\perp}}{\mu'_0}{\cal F}[w_{S_1}, h_{1T}, {\bar h}_1^{\perp}], 
\label{s1p}
\\ 
S'_2 &=& A'(y)\frac{q_{\perp}}{\mu_0}{\cal F}[w_{S_2}, f_{1T}^{\perp}, {\bar f}_1], 
\label{siv}
\\
S'_3 &=& -C'(y)\frac{q_{\perp}^3}{\mu_0^2\mu'_0}{\cal F}[w_{S_3}, h_{1T}^{\perp}, {\bar 
h}_1^{\perp}],\label{s3p} 
\\
A'(y) &=& 1/2-y + y^2,  \ ~~~~~ \ C'(y) =y(1-y) \label{acp}
\end{eqnarray}
and
\begin{equation} 
y = 1/2(1+cos \theta),
\end{equation} 
$\theta$ being the polar angle of the positive lepton in one of the frames (GJ, UC, CS) 
defined at the beginning of this section. $\mu_0$ and $\mu'_0$ are energy scales relative to 
the two initial hadrons in the Drell-Yan process. The change of sign of the T-odd functions 
with respect to SIDIS has been taken into account in the coefficients $S'_1$, $S'_2$ and 
$S'_3$, as already discussed at the end of subsect. 2.3. 

\begin{figure}[htb]
\begin{center}
\epsfig{file=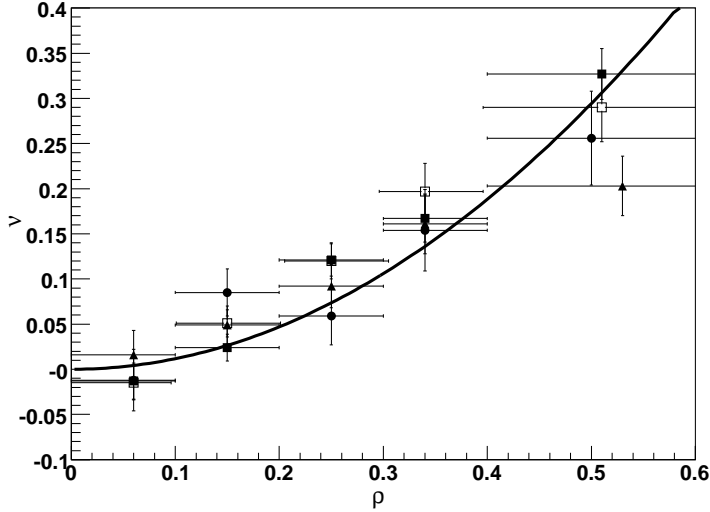,height=7.5cm}
\caption{Behavior of the asymmetry parameter $\nu$ {\it vs} the dimensionless 
parameter $\rho = q_{\perp}/Q$. Data are taken from refs.\cite{fa1,fa2}: circles  correspond 
to $\sqrt{s}$ = 16.2 GeV, squares to $\sqrt{s}$ = 19.1 GeV and triangles to $\sqrt{s}$ = 23.2 
GeV. The best fit is made with formula (\ref{nu1}), $A_0$ = 1.177.} 
\end{center}
\label{fig:one}
\end{figure}

\section{Determining $\mu_0$}

Here we derive the appropriate value of the parameter $\mu_0$ for sufficiently large $Q^2$. 
To this end we expand (see appendix) the correlator in powers of the coupling and, by 
exploiting the Politzer\cite{pol} theorem on equations of motion (see also ref.\cite{qiu}), 
we compare the zero order term and the first order correction respectively with the T-even 
and with the T-odd parameterizations, eqs. (\ref{par0}) and (\ref{cmo}). We shall show that 
the two procedures lead to consistent results. 

\subsection{Spin Density Matrix}

In appendix we show that, in the case of a transversely polarized nucleon, one 
has\cite{fe,dis2}, in the limit of $g \to 0$,
\begin{equation} 
\Phi \to \rho = \frac{1}{2}(\rlap/p+m)[f_1(x, {\bf p}_{\perp}^2)+\gamma_5\rlap/S_q h_{1T}(x, 
{\bf p}_{\perp}^2)]. \label{d2}
\end{equation} 
Here $m$ is the rest mass of the quark, such that $p^2$ = $m^2$, and $S_q$ is (up to a sign) 
the quark Pauli-Lubanski vector, defined so as to coincide, in the {\it quark} rest 
frame\cite{ael}, with the Pauli-Lubanski vector $S$ of the nucleon in its rest frame. 
Now we compare the various Dirac components  of the density matrix (\ref{d2}) with those of 
the T-even correlator (\ref{par0}), taking into account relation (\ref{qspin}) between $S$ 
and $S_q$, and
\begin{equation} 
p = \sqrt{2}{\cal P} n_+ +p_{\perp} +O\left({\cal P}^{-1}\right). \label{rel02}
\end{equation}
As a result we get the following relations for a free, on-shell quark\cite{dis2}: 
\begin{eqnarray} 
\lambda_{\perp} h_{1T}^{\perp} &=& (1-\epsilon_1) {\bar \lambda}_{\perp}h_{1T}, \label{ff1}
\\
\lambda_{\perp} g_{1T} &=& (1-\epsilon_2) {\bar \lambda}_{\perp}h_{1T}.\label{ff5}
\end{eqnarray}
Here 
\begin{equation}
{\bar \lambda}_{\perp} = -p_{\perp}\cdot S/{\cal P}, \label{lchel}
\end{equation}
moreover $\epsilon_1 \simeq m/{\cal P}$ and $\epsilon_2 \simeq m/2{\cal P}$ are the 
correction terms due to the quark mass, which is small for light flavors. The terms of order 
$O\left[(m^2+{\bf p}_{\perp}^2)/{\cal P}^2\right]$ have been neglected.

In order to determine $\mu_0$, we observe that the functions $g_{1T}$, $h_{1T}$ and 
$h_{1T}^{\perp}$, involved in formulae (\ref{ff1}) and (\ref{ff5}), are twist 2, therefore 
they may be interpreted as quark densities. For example, $g_{1T}$ is the helicity density of 
a quark in a tranversely polarized nucleon. Therefore it is natural to fix $\mu_0$ in such a 
way that $g_{1T}$ and $h_{1T}^{\perp}$ are normalized like $h_{1T}$. This implies, neglecting 
the quark mass,
\begin{equation}
\lambda_{\perp} = \bar{\lambda}_{\perp}, \label{lambd}
\end{equation}
and, according to eqs. (\ref{li}) and (\ref{lchel}), 
\begin{equation}
\mu_0 = {\cal P} = \frac{1}{\sqrt{2}}p\cdot n_-. \label{mu0}
\end{equation}

\subsection{First Order Correction in the Coupling}

We show in appendix that the first order correction in $g$ of the correlator - denoted as 
$\Phi_1$ in the following - is T-odd and corresponds to a quark-gluon-quark correlation. This 
confirms that T-odd functions vanish if we neglect quark-gluon interactions inside the 
hadron. Moreover the result binds us to compare $\Phi_1$ to the  parameterization (\ref{cmo}) 
of the T-odd correlator. As proven in appendix, the comparison yields
\begin{equation}
\mu_0 \propto {\cal P}, \label{mu00}
\end{equation} 
consistent with eq. (\ref{mu0}). Result (\ref{mu00}) is a consequence of the Politzer 
theorem, of four-momentum conservation and of kinematics of one-gluon exchange. But the T-odd 
functions $h_1^{\perp}$ and $f_{1T}^{\perp}$ may be interpreted as quark densities (see 
subsect. 2.3), provided they are properly normalized. Therefore we adopt for them the same 
normalization as for the T-even density functions - {\it e. g.}, $g_{1T}$ - which involve the 
vector $\eta_{\perp}$. This leads again to result (\ref{mu0}), analogously to the case of 
noninteracting partons. Therefore we conclude that, for sufficiently large $Q^2$, the energy 
scale $\mu_0$ has to be identified with ${\cal P}$. This is true both for some (T-even) 
functions of the parameterization (\ref{par0}) - which survive also in absence of quark-gluon 
interactions - and for the T-odd correlator (\ref{cmo}), which, on the contrary, depends on 
such interactions. Thus the conjecture proposed at the end of subsect. 2.2 is confirmed. 
Incidentally, eqs. (\ref{ff1}) and (\ref{ff5}) involve just T-even functions, therefore, 
according to the results deduced in appendix, these equations are valid up to terms of 
$O(g^2)$ and are expected to hold down to reasonably small $Q^2$. 

\subsection{Remarks}

At this point some important remarks are in order.

a) Of course, we expect $\mu_0$ to be modified by nonperturbative interactions: for example, 
in the case of the already cited quark-diquark model[35-37, 44-47] (see subsect. 2.3), the 
interference term scales with $Q^2$, in agreement with the assumption $\mu_0 = M$ 
\cite{rs,mt}. However,  the virtuality of the exchanged gluon increases proportionally to 
$Q^2$ \cite{cs2,cs3}, so that, at increasing $Q^2$, the gluon "sees" the single partons 
rather than the diquark as a whole and eq. (\ref{mu0}) appears more appropriate. To 
summarize, if we take into account the intrinsic transverse momentum of quarks, we are faced 
with the {\it normalization} scale $\mu_0$, which, for large $Q^2$, is equal to ${\cal P}$, 
while for smaller $Q^2$ (such that nonperturbative interactions are not negligible) it is of 
the order of the hadron mass, in accord with a phase space restriction.

b) Since, as already observed, the parameter $\mu_0$ is encoded in the four-vector 
$\eta_{\perp} = p_{\perp}/\mu_0$, result (\ref{mu0})  implies that, for sufficiently large 
$Q^2$, the transverse momentum has to be treated as an effective higher twist. This agrees 
with the observation by Qiu and Sterman\cite{qs1,qs2,qs3} that, owing to gauge invariance, 
transverse momentum has to be paired with a transversely polarized gluon, which, through 
quark-gluon-quark correlations, gives rise to a higher twist contribution. But, as already 
stressed, and as shown in appendix, T-odd functions may be viewed as correlations of this 
kind. On the other hand, also the function $g_{1T}$, T-even and classified as twist 2, 
results to be suppressed for large ${\cal P}$. All that casts some doubts on the correlation 
between the twist of an operator and the ${\cal P}$ dependence of the corresponding 
coefficient\cite{te1,te2}, when transverse momentum dependent functions are involved. Indeed, 
in this case, although the Dirac operator commutes with the  Hamiltonian of a free 
quark\cite{jj1,jj2}, its mean value over the nucleon state may be suppressed, if the density 
function describes interference between two quark states with different orbital angular 
momenta. This occurs for those "soft" functions which are washed out by ${\bf 
p}_{\perp}$-integration.  

c) It is worth comparing our approach to Kotzinian's\cite{ko}, who starts from the 
approximate expression of the density matrix for a free ultra-relativistic fermion and adapts 
it to the case of a quark in the nucleon. He parameterizes the density matrix with the 6  
twist-2, T-even functions that appear in the parameterization (\ref{par0}). Similar results 
are obtained by Ralston and Soper\cite{rs} and by Tangerman and Mulders\cite{tm1}. The 
difference with our approach is that those authors do not take into account the Politzer 
theorem\cite{pol}, which implies relations among the "soft" functions. 

\begin{figure}[htb]
\begin{center}
\epsfig{file=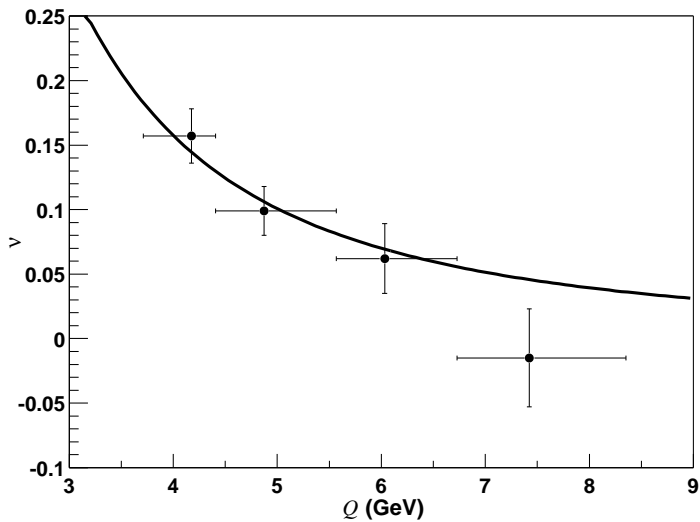,height=7.5cm}
\caption{Behavior of the asymmetry parameter $\nu$ {\it vs} the effective mass $Q$
of the final lepton pair at fixed $q_{\perp}$. $\sqrt{s}$ = 23.2 $GeV$ Data from 
ref.\cite{fa2}, CS frame, and fitted with formula (\ref{nu1}), $A_0\cdot q^2_{\perp}$ = 2.52 
$GeV^2$.}
\end{center}
\label{fig:two}
\end{figure}

\subsection{Determining $\mu_0$ in the Fragmentation Correlator}

As regards the fragmentation correlator, we adapt our previous line of reasoning to the case 
of a quark fragmenting into a transversely polarized spin-1/2 particle, say a $\Lambda$. For 
$g\to 0$ one has
\begin{equation} 
\Delta \to \rho' = \frac{1}{2}(\rlap/k+M')(D+H_1\gamma_5\rlap/S'). \label{df2}
\end{equation}
Here $M'$ and $S'$ are, respectively, the mass and the Pauli-Lubanski vector of the final 
hadron, whereas $H_1$ is the transversely polarized fragmentation function; the other symbols 
are those introduced in subsect. 2.4. By comparing this limiting expression with a  
parameterization of $\Delta$ - analogous to eq. (\ref{par0}) as regards twist-2 terms - , we 
get $\mu_0^{\pi} = k\cdot n'_+/\sqrt{2}$.

\begin{figure}[htb]
\begin{center}
\epsfig{file=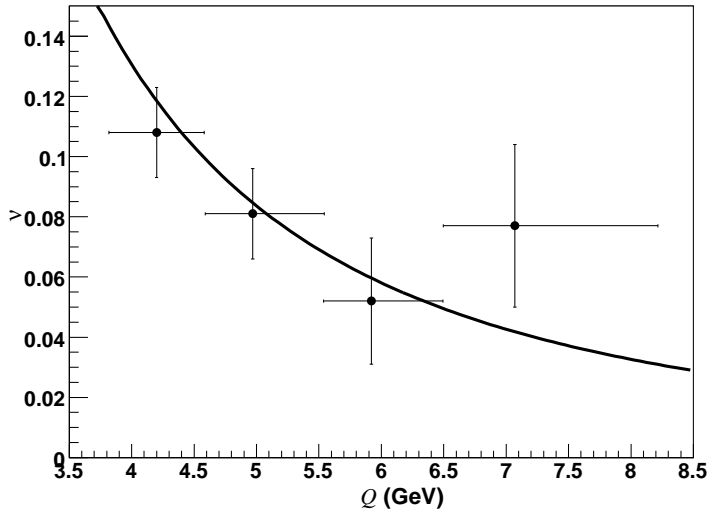,height=7.5cm}
\caption{Same as fig 2. $\sqrt{s}$ = 19.1 $GeV$. $A_0\cdot q^2_{\perp}$ = 2.09 $GeV^2$.}
\end{center}
\label{fig:three}
\end{figure}

\section{${\bf Q^2}$ Dependence of Asymmetries}
Now we apply the result of the previous section to the processes considered in the present 
paper. It is convenient to take the space component of $n_-$ along the direction of one of 
the two initial hadrons (Drell-Yan) or along the direction of the virtual photon (SIDIS). In 
both cases we get ${\cal P}\simeq Q/2$. Therefore we assume
\begin{equation}
\mu_0 = \frac{Q}{2}, \label{mu1}
\end{equation}
the result being trivially extended to $\mu'_0$ and to $\mu_0^{\pi}$. As a consequence, we 
conclude that the azimuthal asymmetries illustrated in sect. 3 decrease with $Q^2$. In 
particular, as regards SIDIS, we predict 
\begin{equation}
S_1, S_2, D_2 \propto \rho, 
\ ~~~~ \ 
U_1 \propto \rho^2, 
\ ~~~~ \
S_3 \propto \rho^3 \label{pv1}
\end{equation}
and
\begin{equation}
D_2  \propto \frac{M}{Q}, 
\ ~~~~ \
S_4 \propto \frac{\rho^2 M}{Q}, \label{pv2}
\end{equation}
where 
\begin{equation}
\rho = q_{\perp}/Q.\label{rho}
\end{equation}
Results (\ref{pv2}) are consequences of the fact that  $\lambda$ [see eq. (\ref{pl})] is 
proportional to $Q^{-1}$ for a transversely polarized nucleon. Such predictions might be 
checked by comparing data of experiments which have been realized 
(HERMES\cite{her1,her2,her3,her4} and COMPASS\cite{co}) with those planned (CLAS\cite{cls}), 
which operate in different ranges of $Q^2$. A strategy could be, for instance, to isolate the 
various Fourier components in the cross section [see eqs. (\ref{u0}) to (\ref{d1})] by means 
of the weighted asymmetries and to study their $Q^2$ dependence. A particular remark is in 
order as regards the unpolarized SIDIS asymmetry, which we predict to decrease as $1/Q^2$, 
just like the twist-4 $cos 2\phi$ asymmetry arising as a consequence of the quark transverse 
momentum\cite{ca1,ca2,ko}. This makes the two asymmetries hardly distinguishable, but the 
last asymmetry can be parameterized, as well as the $cos \phi$ asymmetry (the Cahn 
effect\cite{ans4}), by means of the unpolarized quark density. 

\begin{figure}[htb]
\begin{center}
\epsfig{file=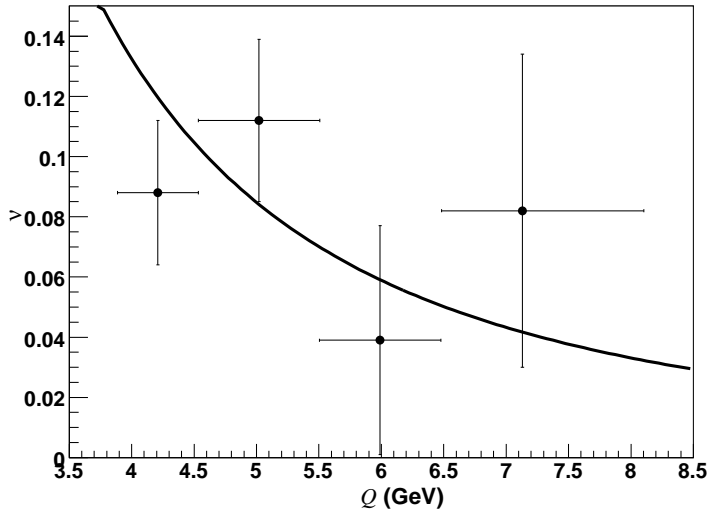,height=7.5cm}
\caption{Same as fig 2. $\sqrt{s}$ = 16.2 $GeV$. $A_0\cdot q^2_{\perp}$ = 2.12 $GeV^2$.}
\end{center}
\label{fig:four}
\end{figure}

Concerning Drell-Yan, the predictions are 
\begin{equation}
S'_1, S'_2 \propto \rho, \label{ssdy}
\ ~~~~ \ 
U'_1 \propto \rho^2, 
\ ~~~~ \
S'_3 \propto \rho^3. \label{pvv}
\end{equation}
As regards $U'_1$, the result will be checked  against unpolarized Drell-Yan data in the next 
section; the other three predictions could be verified, in principle, by comparison with data 
from  experiments planned at various facilities, like RHIC\cite{rh}, GSI\cite{gs1,gs2} and 
FNAL\cite{fn}. It is important to observe that, according to the approximation assumed in the 
present article, the asymmetry terms $S'_1$, $S'_2$, $S'_3$ and $U'_1$ are invariant under 
rotations (see eqs. (\ref{u1p}) to (\ref{s3p})) and therefore independent of the frame chosen 
(CS, GJ or UC).

As a conclusion of this section it is worth recalling that the Drell-Yan single spin 
asymmetry, integrated over the transverse momentum of the final muon pair, was studied some 
years ago, in terms of a quark-gluon-quark correlation function, and it was found to decrease 
as $Q^{-1}$ \cite{bmt,dis4,hts,bm2,bq} (see also refs.[98-102, 124]), consistent with our 
result (\ref{ssdy}). 

\begin{figure}[htb]
\begin{center}
\epsfig{file=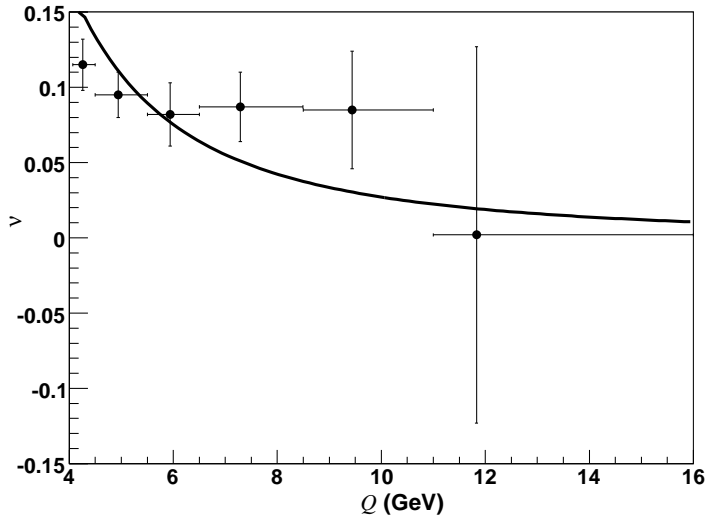,height=7.5cm}
\caption{Same as fig 2. Data from ref.\cite{fa1}, $\sqrt{s}$ = 19.1 $GeV$. $A_0\cdot 
q^2_{\perp}$ = 2.71 $GeV^2$.}
\end{center}
\label{fig:five}
\end{figure}

\section{Azimuthal Asymmetry in Unpolarized Drell-Yan}

 As is well-known, unpolarized Drell-Yan presents an azimuthal asymmetry. This has been seen, 
for example, in reactions of the type\cite{fa1,fa2,fa3}
\begin{equation}
\pi^- N \to \mu^+ \mu^- X, \label{dyan}
\end{equation}
where $N$ is an unpolarized tungsten or deuterium target, which scatters off a negative pion 
beam. The Drell-Yan angular differential cross section is conventionally expressed as
\begin{equation}
\frac{1}{\sigma}\frac{d\sigma}{d\Omega} = \frac{3}{4\pi} \frac{1}{\lambda+3}
\left(1+\lambda cos^{2}\theta +\mu sin 2\theta cos\phi+\frac{1}{2} \nu sin^2\theta cos 
2\phi\right). \label{dycs}
\end{equation}
Here $\Omega$ = ($\theta$, $\phi$), $\theta$ and $\phi$ being respectively the polar and 
azimuthal angle of the $\mu^+$ momentum in the one of the frames defined in sect. 3. Moreover 
$\lambda$, $\mu$ and $\nu$ are parameters, which are functions of the overall center-of-mass 
energy squared $s$, of  $q_{\perp}^2$, of $Q$ and of the Feynman longitudinal fractional 
momentum $x_F$ of the muon pair with respect to the initial beam. 

In the naive Drell-Yan model, where the parton transverse momentum and QCD corrections are 
neglected, one has $\lambda$ = 1, $\mu$ = $\nu$ = 0. Therefore deviations of such parameters 
from the above predictions - observed experimentally both for $\lambda$ and $\nu$, while 
$\mu$ is consistent with 0 \cite{fa1,fa2,fa3} - can be attributed to transverse momentum or 
gluon effects, as illustrated in ref.\cite{fa1}. The main contribution to the Drell-Yan cross 
section derives from QCD first order effects, typically from the partonic reactions $q{\bar 
q}\to g\gamma^*$ and $q g \to q\gamma^*$ \cite{fa1,fa2,fa3,sv}, which also could account for 
the asymmetry parameter $\nu$ \cite{fa1,fa2,fa3,bv}. However, such effects fulfill the 
Lam-Tung\cite{ltu} relation, which, instead, turns out to be rather strongly violated (see 
refs.\cite{bbh,fa1,fa2,fa3} and refs. therein). This fact has led people to propose 
alternative mechanisms\cite{bbh,bbnu} for explaining the asymmetry. Furthermore we notice 
that data\cite{fa1,fa2,fa3} exhibit for $\nu$ a substantial independence of the frame chosen 
(CS, GJ, UC), just as predicted by T-odd functions (see the previous section), while first 
order perturbative QCD corrections would imply\cite{bv} a considerable frame dependence for 
that parameter.

The behavior of Drell-Yan data may be understood by observing that the cross section is very 
sensitive also to power corrections\cite{llm,bb1,bb2} (see also refs.\cite{be1,be2}). In 
particular, a $\lambda \neq 1$ is obtained by assuming for reaction (\ref{dyan}) a simple 
model of initial state interactions\cite{bb1,bb2}, somewhat similar to the quark-gluon-quark 
correlations\cite{qs1,qs2,qs3}. Here the Drell-Yan unpolarized cross section is of the type
\begin{equation} 
d\sigma \propto |f_0+f_1|^2, \label{bbb}
\end{equation}
where  $f_0$ is the naive Drell-Yan amplitude and $f_1$ consists of two terms (due to gauge 
invariance), describing one gluon exchange between the spectator quark of the meson and each 
active parton. It results\cite{bb1,bb2} $|f_0|^2\propto (1-x)^2(1+cos^2\theta)$,
$|f_1|^2\propto \rho^2cos^2\theta$ and $2\Re f_0f_1^*\propto (1-x)\rho sin^2\theta cos\phi 
cos\psi_0$, where $\psi_0$ is the relative phase of the two amplitudes. This implies 
$\lambda-1$ $\propto$ $\rho^2/(1-x)^2$, in good agreement with data\cite{fa1,fa2,fa3}. 
However, $\mu$ depends crucially on $\psi_0$, moreover the third term of eq. (\ref{dycs}) is 
absent. This could be recovered by inserting in eq. (\ref{bbb}) a third amplitude, say 
$f'_1$, describing one gluon exchange between each active parton and the spectator partons of 
the nucleon: the missing asymmetry is reproduced by the interference term $2\Re f'_1f_1^*$, 
as sketched at the end of subsect. 2.3, which amounts to recovering T-odd 
functions\cite{bbh}.

Indeed, the features of the parameters $\mu$ and $\nu$ are suitably interpreted in terms of 
the correlator: comparison of eq. (\ref{dyu}) with eq. (\ref{dycs}) yields $\mu = 0$ and   
\begin{equation}
\nu = A_0 \frac{q_{\perp}^2}{Q^2} = A_0 \rho^2, \label{nu1}
\end{equation}
with 
\begin{equation}
A_0 = \frac{{\cal F}[w'_{U_1}, h_1^{\perp}, {\bar h}_1^{\perp}]}{{\cal F}[w'_{U_0}, f_1, 
{\bar f}_1]}.  \label{nu2}
\end{equation}
Here eqs. (\ref{u0p}), (\ref{u1p}), (\ref{acp}) and the second eq. (\ref{pvv}) have been 
taken into account. We make some approximations concerning $A_0$. First of all, we neglect 
its $q_{\perp}$ dependence: for example, if we assume a gaussian behavior as regards the 
${\bf p}_{\perp}$-dependence of the density functions, the $q_{\perp}$ dependence disappears 
in the ratio (\ref{nu2}). Secondly we neglect the $Q^2$ evolution of the "soft" functions 
involved; such a dependence is expected to be quite smooth, as follows by assuming 
factorization and demanding factorization scale independence for the hadronic 
tensor\cite{st}. Lastly, as told in subsect. 2.1, we neglect the Sudakov suppression: indeed, 
this effect, as well as the previous one, would imply a weak $Q^2$ dependence\cite{bo0,bo1}, 
more complicated than the (approximate) $\rho$-dependence\cite{bo0,bo1} exhibited by data 
(see fig. 1 of the present paper and tables 1 to 3 in ref.\cite{fa1}). Therefore we 
approximate $A_0$ by a constant.  

We fit formula (\ref{nu1}) to the experimental results of $\nu$ at different energies, both 
as a function of $\rho$ (fig. 1) and as a function of $Q$ at fixed $q_{\perp}$ (figs. 2 to 
6), treating $A_0$ as a free parameter. We stress that the $\rho$-dependence of $\nu$ cannot 
be reproduced by the assumption $\mu_0 = M$ \cite{tm1,tm2,mt}, not even taking into account 
the effects, just discussed, of QCD evolution and Sudakov suppression. This assumption would 
provide also a poor approximation to data of $\nu$ {\it versus} $Q$ at fixed $q_{\perp}$.  

\begin{figure}[htb]
\begin{center}
\epsfig{file=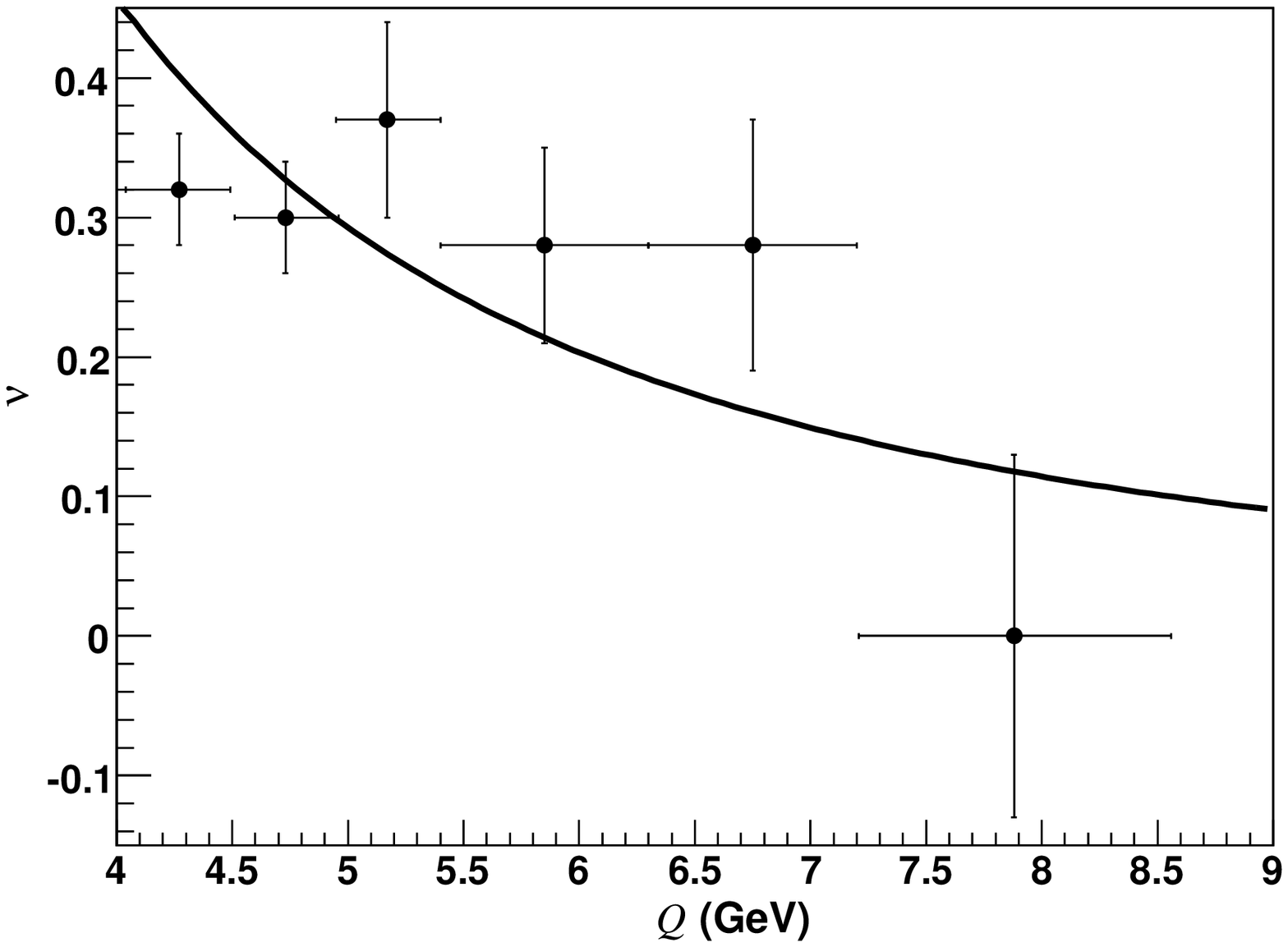,height=7.5cm}
\caption{Same as fig 2. Data from ref.\cite{fa3}, $\sqrt{s}$ = 21.7 $GeV$. $A_0\cdot 
q^2_{\perp}$ = 7.32 $GeV^2$.}
\end{center}
\label{fig:six}
\end{figure}

\section{Conclusion}

We have studied the parameterization of the transverse momentum dependent quark correlator, 
both for distributions inside the hadron and for fragmentation processes. We are faced with 
the energy scale $\mu_0$, introduced in the parameterization for dimensional reasons, and 
determining the normalization of some of the quark densities (or fragmentation functions) 
involved. Comparison of the parameterization with the limiting expression of the correlator 
for noninteracting quarks yields $\mu_0 = p\cdot n_-$, contrary to the usual\cite{tm1,tm2,mt} 
assumption, $\mu_0 = M$, which appears more appropriate for situations where nonperturbative 
interactions among partons are present. The two different assumptions lead to different 
predictions on the $Q^2$ dependence of azimuthal asymmetries in SIDIS and Drell-Yan. Our 
result agrees with previous approaches to azimuthal asymmetries, in particular with the $Q^2$ 
dependence predicted by quark-quark-gluon correlations\cite{qs1,qs2,qs3} for Drell-Yan single 
spin 
asymmetry, and also with data of azimuthal asymmetry in unpolarized Drell-Yan. In particular, 
our interpretation of this asymmetry is considerably simpler than the one which could be 
obtained with the usual assumption about $\mu_0$. Further challenges to the two different 
theoretical predictions could come from future Drell-Yan experiments\cite{rh,gs1,gs2,fn} and 
from comparison between present\cite{her1,her2,her3,her4,co} and incoming\cite{cls} SIDIS 
data.

\vspace {10pt}

\centerline{\bf Acknowledgments}
The author is deeply indebted to his friend A. Di Giacomo for useful suggestions and 
constructive criticism. He also thanks profs. S. Brodsky and D. Sivers for acute and 
stimulating observations.

\vspace {10pt}

\setcounter{equation}{0}
\renewcommand\theequation{A. \arabic{equation}}

\appendix{\large \bf Appendix}

We illustrate some features of the correlator; in particular, we expand this quantity in 
powers of the coupling and study in detail the zero and first order term of the expansion. 
The correlator is defined as\cite{mt}
\begin{equation}
\Phi = N \int \Phi' (p; P,S)dp^-.\label{corr}
\end{equation}
Here $N$ is a normalization constant, to be determined below, and $\Phi' (p; P,S)$ is defined 
in such a way that
\begin{equation}
\Phi'_{ij}(p; P,S) = \int\frac{d^4x}{(2\pi)^4} e^{ipx} 
\langle P,S|\bar{\psi}_j(0) {\cal L}(x)  \psi_i(x)|P,S\rangle, \label{corr1}
\end{equation}
$\psi$ being the quark (or antiquark) field of a given flavor and $|P,S\rangle$ a state of a 
nucleon with a given four-momentum $P$ and Pauli-Lubanski four-vector $S$, while $p$ is the 
quark four-momentum. The color index has been omitted in $\psi$ for the sake of simplicity. 
Moreover
\begin{equation}
{\cal L}(x) = {\mathrm P} exp\left[-ig\Lambda_{\cal I}(x)\right], \ ~~~~~ \ {\mathrm with}
\ ~~~~~ \ \Lambda_{\cal I}(x) = \int_{0({\cal I})}^x \lambda_a A^a_{\mu}(z)dz^{\mu}, 
\label{link}
\end{equation}
is the gauge link operator. Here $g$ is the coupling and "P" denotes the path-ordered product 
along a given integration contour ${\cal I}$, $\lambda_a$ and $A^a_{\mu}$ being respectively 
the Gell-Mann matrices and the gluon fields. The link operator depends on the choice of 
${\cal I}$, which has to be fixed so as to make a physical sense. According to previous 
treatments\cite{co2,mt}, we define two different contours, ${\cal I}_{\pm}$, as sets of three 
pieces of straight lines, from the origin to $x_{1\infty}\equiv (\pm\infty, 0, {\bf 
0}_{\perp})$, from $x_{1\infty}$ to $x_{2\infty}\equiv (\pm\infty, x^+, {\bf x}_{\perp})$ and 
from $x_{2\infty}$ to  $x\equiv (x^-, x^+,{\bf x}_{\perp})$; here the $\pm$ sign has to be 
chosen, according as to whether final or initial state interactions\cite{co2,mt} are involved 
in the reaction. We have adopted a frame - to be used throughout this appendix - whose 
$z$-axis is taken along the nucleon momentum, with $x^{\pm} = 1/\sqrt{2}(t\pm z)$. 
\vskip 0.30in

{\bf ~~ T-even and T-odd correlator}

We set
\begin{equation}
\Phi'_{E(O)} = \frac{1}{2}[\Phi'_+\pm\Phi'_-], \label{spl}
\end{equation}
where $\Phi'_{\pm}$ corresponds to the contour ${\cal I}_{\pm}$ in eqs. (\ref{link}), while
$\Phi'_E$ and $\Phi'_O$ select respectively the T-even and the T-odd "soft" functions. These 
two correlators contain respectively the link operators ${\cal L}_E(x)$ and ${\cal L}_O(x)$, 
where
\begin{equation}
{\cal L}_{E(O)}(x) = \frac{1}{2} {\mathrm P} \left\{exp\left[-ig\Lambda_{{\cal 
I}_+}(x)\right]\pm 
exp\left[-ig\Lambda_{{\cal I}_-}(x)\right]\right\} \label{spl1}
\end{equation}
and $\Lambda_{{\cal I}_{\pm}}(x)$ are defined by the second eq. (\ref{link}).  It is 
convenient to consider an axial gauge with antisymmetric boundary conditions\cite{mt}, to be 
named $G$-gauge in the following. This yields 
\begin{equation}
\Lambda_{{\cal I}_-}(x) = -\Lambda_{{\cal I}_+}(x)\label{link0}
\end{equation}
and therefore
\begin{equation}
{\cal L}_E(x) = {\mathrm P} cos\left[g\Lambda_{{\cal I}_+}(x)\right], \ ~~~~~~ \
{\cal L}_O(x) = -i{\mathrm P} sin\left[g\Lambda_{{\cal I}_+}(x)\right].
\end{equation}
Then the T-even (T-odd) part of the correlator consists of a series of even (odd) powers of 
$g$, each term being endowed with an even (odd) number of gluon legs. Moreover eq. 
(\ref{link0}) implies that the T-even functions are independent of the contour (${\cal I}_+$ 
or ${\cal I}_-$), while T-odd ones change sign according as to whether they are generated by 
initial or final state interactions\cite{co2}. In this sense, such functions are not strictly 
universal\cite{co2}. 

The two conclusions above, as well as the power expansion in the coupling, turn out to be 
gauge independent, since the the correlator is by definition gauge independent for {\it any} 
value of $g$ and the same is true for any term in the expansion. As a consequence, the zero 
order term is T-even, while the first order correction is T-odd. This confirms that no T-odd 
terms occur without interactions among partons, as claimed also by other 
authors\cite{bhs1,bhs2,bhs3,co2}. 

Let us consider the expansion of $\Phi'$ in powers of $g$, {\it i. e.,}
\begin{equation}
\Phi' = \Phi'_0 - ig \Phi'_1 + ..., \label{exp}
\end{equation} 
with
\begin{equation}
(\Phi'_{0})_{ij} = \int\frac{d^4x}{(2\pi)^4} e^{ipx} 
\langle P,S|\bar{\psi}_j(0) \psi_i(x)|P,S\rangle \label{cor0}
\end{equation} 
and
\begin{equation}
(\Phi'_{1})_{ij} = \int\frac{d^4x}{(2\pi)^4} e^{ipx} \int_{0({\cal I})}^x dz^{\mu}
\langle P,S|\bar{\psi}_j(0) \lambda_a A^a_{\mu}(z) \psi_i(x)|P,S\rangle. \label{cor1} 
\end{equation}
We stress that the term (\ref{cor1}) consists of a quark-gluon-quark correlation, analogous 
to the one introduced by Efremov and Teryaev\cite{et2,et3} and by Qiu and 
Sterman\cite{qs1,qs2,qs3}. Inserting expansion (\ref{exp}) into eq. (\ref{corr}), we get 
\begin{equation}
\Phi = \Phi_0 + \Phi_1 + ..., \label{exp1}
\end{equation} 
where
\begin{equation}
\Phi_0 = N \int \Phi'_0 (p; P,S)dp^- ~~~~ {\mathrm and} ~~~~ 
\Phi_1 = -i g N \int \Phi'_1 (p; P,S)dp^-. \label{corap}
\end{equation}
From now on we shall consider a {\it transversely} polarized nucleon. Then, according to our 
previous considerations, we may identify, at the two lowest orders in $g$, $\Phi_0$ with 
$\Phi_e$ and $\Phi_1$ with $\Phi_o$, where $\Phi_e$ and $\Phi_o$ are given, respectively, by 
eqs. (\ref{par0}) and (\ref{cmo}) in the text. Now we study in detail these two terms of the 
expansion. 

\vskip 0.30in
{\bf ~~ Zero order term}

We apply the Politzer theorem on equations of motion\cite{pol}, {\it i.e.}, 
\begin{equation}
\langle P,S|{\bar \psi(0)}{\cal L}(x) (iD\hspace{-0.7em}/-m)\psi(x)|P,S\rangle = 0. 
\label{pol}
\end{equation}
Here $D_{\mu} = \partial_{\mu} - ig \lambda_a A^a_{\mu}$ is the covariant derivative. The 
result (\ref{pol}) survives renormalization and applies also to off-shell quarks. Expanding 
${\cal L}(x)$ in powers of $g$, at zero order the theorem implies that the quark can be 
treated as if it were on shell (see also ref.\cite{qiu}). Then we consider the Fourier 
expansion of the unrenormalized field of an on-shell quark, {\it i. e.},
\begin{equation} 
\psi(x) = \int \frac{d^4p}{(2\pi)^{3/2}} \sqrt{\frac{m}{\cal P}} 
\delta\left(p^--\frac{\sqrt{m^2+{\bf p}^2_{\perp}}}{2p^+}\right) e^{-ipx} \sum_s u_s(p) 
c_s(p), \label{field}
\end{equation} 
where $m$ is the rest mass of the quark, $s = \pm 1/2$ its spin component along the nucleon 
polarization in the quark rest frame, $u$ its four-spinor, $c$ the destruction operator for 
the flavor considered and ${\cal P}$ is defined by the first eq. (\ref{li}) in the text: in 
our frame ${\cal P} = p^+/\sqrt{2}$. As regards the normalization of $u_s$ and $c_s$, we 
assume
\begin{equation} 
{\bar u}_s u_s = 2m, ~~~~~~~~ \ \langle P,S|c_s^{\dagger}({\tilde {\bf p}'})c_{s'}({\tilde 
{\bf p}})|P,S\rangle = (2\pi)^3\delta^3({\tilde {\bf p}'}-{\tilde {\bf p}}) \delta_{ss'} 
q_s({\tilde {\bf p}}), \label{norm}
\end{equation}
where ${\tilde {\bf p}}\equiv(p^+,{\bf p}_{\perp})$ and $q_s({\tilde {\bf p}})$ is the 
probability density to find a quark with spin component $s$ along the nucleon (transverse) 
polarization and four-momentum $p \equiv (p^-, {\tilde {\bf p}})$, with $p^- = 
{\sqrt{m^2+{\bf p}^2_{\perp}}}/{2p^+}$. For an antiquark the definition is quite analogous. 
Substituting eq. (\ref{field}) and the second eq. (\ref{norm}) into eq. (\ref{cor0}) and into 
the first eq. (\ref{corap}), we get
\begin{equation}
(\Phi_0)_{ij} = \frac{N}{2{\cal P}} \sum_{s} q_s({\tilde {\bf p}}) [u_{s}({\tilde {\bf 
p}})]_i[\bar{u}_{s}({\tilde {\bf p}})]_j. \label{dm2}
\end{equation}
But $[u_{s}({\tilde {\bf p}})]_i[\bar{u}_{s}({\tilde {\bf p}})]_j$ is nothing but the matrix 
element $\rho_{ij}$ of the spin density matrix of the quark. Therefore, taking into account 
the first eq. (\ref{norm}), eq. (\ref{dm2}) yields
\begin{equation} 
\Phi_0 = \frac{N}{2{\cal P}}\sum_s q_s ({\tilde {\bf p}})\frac{1}{2}(\rlap/p+m) 
(1+2s\gamma_5\rlap/S_q). \label{dd22} 
\end{equation} 
where $2s S_q$ is the Pauli-Lubanski vector of the quark in a transversely polarized nucleon.
We normalize the correlator according to the definition (\ref{ht}) of the hadronic tensor, 
that is, demanding that it reduce to the spin density matrix in the limit of $g\to 0$. 
Therefore
\begin{equation} 
N = 2{\cal P}. \label{norm1}
\end{equation} 
Eq. (\ref{dd22}) can be conveniently rewritten as
\begin{equation} 
\Phi_0 = \frac{1}{2}(\rlap/p+m)[f_1(x, {\bf p}_{\perp}^2)+\gamma_5\rlap/S_q h_{1T}(x, {\bf 
p}_{\perp}^2)], \label{ddd2}
\end{equation} 
where we have set, according to the definitions of the density functions [in the scaling 
limit, $q_s({\tilde {\bf p}})\to q_s(x,{\bf p}^2_{\perp})$], 
\begin{equation}
f_1 = \sum_{s=\pm 1/2} q_s, \ ~~~~~~ \ h_{1T} = \sum_{s=\pm 1/2} 2s q_s. \label{dens33}
\end{equation}
Eq. (\ref{ddd2}) corresponds to formula (\ref{d2}) in the text. According to the Politzer 
theorem, renormalization modifies the functions $f_1$ and $h_{1T}$, but not the structure of 
this expression.

Now we express $S_q$ as a function of $S$. The two vectors do not coincide, since the spin 
operator for a massive particle has to be defined in the particle rest frame\cite{ael}. 
Taking into account the proper Lorentz boosts, we get, for a transversely polarized nucleon,
\begin{equation}
S^q = S + \bar{\lambda}_{\perp}\frac{p}{m} + O(\bar{\eta}^2_{\perp}),  \label{qspin}
\end{equation}
with $\bar{\eta}_{\perp} = p_{\perp}/{\cal P}$ and $\bar{\lambda}_{\perp} = 
-S\cdot\bar{\eta}_{\perp}$.
\vskip 0.30in
{\bf ~~ First order correction}

Now we consider the term (\ref{cor1}), which, as shown before, is T-odd. This term gives rise 
to a final (or initial) state interaction\cite{bmt,bam,bmp,hts,bhs1,bhs2,bhs3}, in the sense 
that the spectator partons of a given hadron may exchange a gluon either with the final 
active quark (in SIDIS) or with the initial active quark of another hadron (in Drell-Yan). 
This kind of interaction selects the direction of the momentum of the gluon in the triple 
correlation (\ref{cor1}). Indeed, if the gluon is emitted by the spectator partons, it must 
have the 
same direction as the active quarks in that correlation; if absorbed, it must have opposite 
direction. This observation is quite important, as we shall see. 

We apply again the Politzer theorem, eq. (\ref{pol}), now considering the first order 
correction of ${\cal L}(x)$. We get, adopting the $G$-gauge,

\begin{equation}
(p\hspace{-0.45em}/-m)\Phi'_1 = {\cal M}, \label{odd1}
\end{equation}
where 
\begin{equation}
{\cal M}_{ij} = \int\frac{d^4x}{(2\pi)^4} e^{ipx} \langle P,S|\bar{\psi}_j(0) 
[{A\hspace{-0.60em}/}_2-{A\hspace{-0.60em}/}]_{ik} \psi_k(x)|P,S\rangle \label{oddp}
\end{equation}
and ${A\hspace{-0.60em}/} = \lambda_a {A\hspace{-0.60em}/}^a(x)$, ${A\hspace{-0.60em}/}_2 = 
\lambda_a {A\hspace{-0.60em}/}^a(x_{2\infty})$. The matrix ${\cal M}$ is a  quark-gluon-quark 
correlator, with the dimensions of a momentum; this matrix is a "soft" quantity, therefore 
independent of the "hard" scale $p^+$. Eq. (\ref{odd1}) implies 
\begin{equation}
\Phi'_1 =  \frac{p\hspace{-0.45em}/+m}{p^2-m^2}{\cal M}, \label{odd2}
\end{equation}
with $p^2\neq m^2$. On the other hand, by considering the Fourier expansion of the gluon 
field in eq. (\ref{oddp}), and by applying again Politzer's theorem, we conclude that the 
other quark in the triple correlation is on shell. Now we show that
\begin{equation}
p^2\propto (p^+)^2 \label{cond2}
\end{equation}
for $p^+\to\infty$, so that $\Phi'_1$ decreases like $(p^+)^{-1}$ in that limit.

Our previous considerations and four-momentum conservation imply that the quarks in the 
triple correlation have four-momenta $p$ and $p\pm k$ respectively, such that 
\begin{equation}
(p\pm k)^2 =m^2.\label{cond4}
\end{equation}
Here $k$ is the four-momentum of the gluon in the triple correlation, the $\pm$ sign 
referring respectively to gluon emission and absorption. Since the gluon is exchanged between 
two color charges, it is space-like, moreover 
\begin{equation}
k \equiv (k_0, {\bf k}), ~~~ {\mathrm with} ~~~ 0<k_0 \sim |k_x|\sim |k_y| << |k_z| = O(p^+). 
\label{cond3}
\end{equation}
But, according to the previous discussion, the kinematics of one gluon exchange demands $k_z$ 
to be positive for emission and negative for absorption. Then condition (\ref{cond2}) follows 
from eqs. (\ref{cond3}) and (\ref{cond4}). 

However, in order to justify the parameterization (\ref{cmo}) for $\Phi_o(p)$ in the text, 
one has to make an approximation. Indeed, the hadronic tensor in SIDIS and in Drell-Yan is 
not rigorously factorizable into two terms, if we include the mechanism of one gluon exchange 
illustrated before. As discussed by Collins in ref.\cite{co2} and papers therein, an 
approximate factorization may be assumed, provided the transverse momentum of the final 
hadron (in SIDIS) or of the final pair (in Drell-Yan) is much smaller than the "hard" scale, 
see condition (\ref{ineq}). Under such a condition, taking into account results (\ref{odd2}) 
and (\ref{cond2}) and the T-odd character of $\Phi'_1$, this term may be parameterized in the 
following way:
\begin{equation}
-ig\Phi'_1 = \frac{K}{p^+}\left\{{\tilde f}^{\perp}_{1T} 
\epsilon_{\mu\nu\rho\sigma}\gamma^{\mu}n_+^{\nu}p_{\perp}^{\rho}S_{\perp}^{\sigma}+
i{\tilde h}^{\perp}_1\frac{1}{2}[\rlap/p_{\perp},\rlap/n_+]\right\}. \label{paro}
\end{equation}
Here $K$ is a numerical constant and ${\tilde f}^{\perp}_{1T}$ and ${\tilde h}^{\perp}_1$ are 
two "soft" functions. Inserting eq. (\ref{paro}) into the second eq. (\ref{corap}) yields a 
parameterization for $\Phi_1$, which, by comparison to eq. (\ref{cmo}), leads us to conclude 
that $\mu_0\propto {\cal P}$. As discussed in subsect. 4.2, a suitable choice of the 
normalization of the two T-odd "soft" functions (which uniquely fixes the constant $K$) leads 
to eq. (\ref{mu0}), as in the noninteracting case.

\end{document}